\documentclass[prl,twocolumn,
tightenlines,superscriptaddress,nofootinbib,
showpacs]{revtex4}
\usepackage{amssymb,latexsym}
\usepackage{amsmath,amsbsy,bbm}
\usepackage{epsfig,bm}
\usepackage{graphicx,comment}
\usepackage{color}
\DeclareMathOperator{\tr}{tr}

\begin{document}
\def\a{{\alpha}}
\def\b{{\beta}}
\def\d{{\delta}}
\def\D{{\Delta}}
\def\e{{\varepsilon}}
\def\g{{\gamma}}
\def\G{{\Gamma}}
\def\k{{\kappa}}
\def\l{{\lambda}}
\def\L{{\Lambda}}
\def\m{{\mu}}
\def\n{{\nu}}
\def\o{{\omega}}
\def\O{{\Omega}}
\def\S{{\Sigma}}
\def\s{{\sigma}}
\def\th{{\theta}}
\def\X{{\Xi}}

\newcommand{\mnod}{\stackrel{\circ}{M}}

\def\ol#1{{\overline{#1}}}

\def\Dslash{\ol D\hskip-0.65em /}
\def\Dslashe{D\hskip-0.65em /}
\def\Pslash{\ol P\hskip-0.65em /}
\def\lslash{l\hskip-0.35em /}
\def\Pslashe{P\hskip-0.65em /}
\def\Dtslash{\tilde{D} \hskip-0.65em /}

\def\CPT{{$\chi$PT}}
\def\QCPT{{Q$\chi$PT}}
\def\PQCPT{{PQ$\chi$PT}}
\def\tr{\text{tr}}
\def\str{\text{str}}
\def\diag{\text{diag}}
\def\order{{\mathcal O}}

\def\cF{{\mathcal F}}
\def\cC{{\mathcal C}}
\def\cB{{\mathcal B}}
\def\cT{{\mathcal T}}
\def\cQ{{\mathcal Q}}
\def\cL{{\mathcal L}}
\def\cO{{\mathcal O}}
\def\cA{{\mathcal A}}
\def\cQ{{\mathcal Q}}
\def\cR{{\mathcal R}}
\def\cH{{\mathcal H}}
\def\cW{{\mathcal W}}
\def\cM{{\mathcal M}}
\def\cD{{\mathcal D}}
\def\cN{{\mathcal N}}
\def\cP{{\mathcal P}}
\def\cK{{\mathcal K}}
\def\Qt{{\tilde{Q}}}
\def\Dt{{\tilde{D}}}
\def\St{{\tilde{\Sigma}}}
\def\cBt{{\tilde{\mathcal{B}}}}
\def\cDt{{\tilde{\mathcal{D}}}}
\def\cTt{{\tilde{\mathcal{T}}}}
\def\cMt{{\tilde{\mathcal{M}}}}
\def\At{{\tilde{A}}}
\def\cNt{{\tilde{\mathcal{N}}}}
\def\cOt{{\tilde{\mathcal{O}}}}
\def\cPt{{\tilde{\mathcal{P}}}}
\def\cI{{\mathcal{I}}}
\def\cJ{{\mathcal{J}}}
\def\cb{{\cal B}}
\def\cbb{{\overline{\cal B}}}
\def\ct{{\cal T}}
\def\ctt{{\overline{\cal T}}}

\def\eqref#1{{(\ref{#1})}}

 
\title{Hyperon Axial Charges in Two-Flavor Chiral Perturbation Theory }

\author{Fu-Jiun~Jiang}
\email[]{fjjiang@itp.unibe.ch}
\affiliation{Institute for Theoretical Physics, Bern University, Sidlerstrasse 5, CH-3012 Bern, Switzerland}

\author{Brian~C.~Tiburzi}
\email[]{bctiburz@umd.edu}
\affiliation{Maryland Center for Fundamental Physics, Department of Physics, University of Maryland, College Park, MD 20742-4111, USA}

\date{\today}


\begin{abstract}
We use two-flavor heavy baryon chiral perturbation theory to investigate the isovector axial charges of the spin one-half hyperons.
Expressions for these hyperon axial charges are derived at next-to-leading order in the chiral expansion. 
We utilize phenomenological and lattice QCD inputs to assess the convergence of the two-flavor theory,
which appears to be best for cascades. 
\end{abstract}

\maketitle


{\bf Introduction}.
The low-energy structure of hadrons is notoriously difficult to describe quantitatively,
because the underlying QCD dynamics is non-perturbative.
Fortunately numerical simulation of QCD on Euclidean spacetime lattices allows first principles study of hadrons%
~\cite{DeGrand:2006aa}. 
Impressive strides continue to be made, 
as algorithmic advances in conjunction with expanded computing resources have put lattice methods in reach of physical predictions.
Chiral perturbation theory (\CPT)~\cite{Gasser:1983yg,Gasser:1984gg} continues to aid the extraction of physical quantities from lattice QCD.
\CPT\
is an effective low-energy description of QCD based on symmetries, 
and allows for quark mass and lattice volume dependence 
of certain low-energy QCD observables to be addressed in a model-independent fashion.

An effective description of low-energy QCD is possible provided a 
systematic power counting can be established to order the infinite terms in the \CPT\ Lagrangian. 
An 
$SU(2)$ 
chiral expansion is better suited for this task compared to 
$SU(3)$, 
because the eta mass squared, 
$m_\eta^2$, 
is not particularly small compared to the square of the chiral symmetry breaking scale, 
$\L_\chi^2$. 
The inclusion of baryons can be done systematically by treating the baryon mass, 
$M_B$,  
as a large parameter%
~\cite{Jenkins:1990jv,Jenkins:1991es}.
An 
$SU(2)$ 
expansion for baryons is expected to be more effective than 
$SU(3)$, 
because the latter expansion contains terms that scale linearly with  
$m_\eta / M_B \sim 0.5$. 
Recent lattice QCD results, 
moreover, 
empirically indicate trouble with 
$SU(3)$ 
expansions%
~\cite{Allton:2008pn,WalkerLoud:2008bp,Aoki:2008sm}.
For the non-strange hadrons, 
$SU(2)$
\CPT\ has long been utilized; 
while for strange hadrons,
two-flavor expansions have been rather limited until recently%
~\cite{Roessl:1999iu,Frink:2002ht,Beane:2003yx,Tiburzi:2008bk,Flynn:2008tg}.

In this work, 
we compute the isovector axial charges of hyperons
utilizing $SU(2)$ heavy baryon \CPT. 
Phenomenologically these charges enter the 
$p$-wave amplitudes 
for non-leptonic weak decays of hyperons, 
where a long-standing puzzle persists.  
These charges also play an important role in the convergence of \CPT,
because the pion-baryon loop diagrams are generated from the axial couplings in \CPT. 
Recent fully dynamical lattice QCD calculations of the nucleon axial charge%
~\cite{Edwards:2005ym,Yamazaki:2008py}
have been accompanied by the first calculation of hyperon axial charges%
~\cite{Lin:2007ap}.
Comparing the axial couplings obtained, 
$g_A \sim 1.2$, 
$g_{\S \S} \sim 0.8$, 
and 
$g_{\X \X} \sim 0.2$,
suggests that the convergence of \CPT\ will improve with increasing strangeness quantum number.
Using phenomenological and lattice QCD inputs,
we verify this pattern of convergence for the axial charges. 
Furthermore, 
we find that the empirical quark mass extrapolation used in~\cite{Lin:2007ap}
is consistent with our $SU(2)$ formula for the cascade
(and possibly for the sigma).
Thus we expect \CPT\ can be used for a controlled extrapolation of data at smaller pion masses. 
For the nucleon, the situation is less clear.

Our presentation begins with
a brief description of the relevant axial matrix elements defining the isovector hyperon axial charges. 
We map the axial-vector current operator onto $SU(2)$ \CPT\ up to next-to-leading order (NLO)
using the power counting of~\cite{Tiburzi:2008bk}, 
and compute axial charges to that order.
Following this, 
we investigate the size of chiral corrections to hyperon axial charges, 
with inputs from experiment and lattice QCD.
A discussion at the end concludes our work.


{\bf Hyperon axial charges}.
The isovector axial charges of 
spin-$\frac{1}{2}$ 
hyperons are defined from axial-vector matrix elements. 
For the isodoublet of quark fields,
$Q = (u, d )^T$, 
the isospin-raising axial current in QCD is written as 
$J^{+}_{\mu,5} = \ol Q \gamma_{\mu}\gamma_{5}\tau^{+}Q$,
where $\tau^+ = \tau^1 + i \tau^2$ is the usual isospin matrix. 
The axial charges 
$G_{\S\S}$, 
$G_{\X\X}$, 
and 
the charge of the axial transition
$G_{\L\S}$
are defined from the rest-frame matrix elements
\begin{eqnarray}
\langle \Sigma^{0} (\bm{0}) | J^{+}_{\mu,5}| \Sigma^{-} (\bm{0}) \rangle 
&=&  
\frac{1}{\sqrt{2}}
G_{\Sigma\Sigma} \,\,
\ol u_{\Sigma} (\bm{0}) \gamma_\mu \gamma_5 u_{\Sigma} (\bm{0})
,\nonumber \\
\langle \Lambda  (\bm{0}) | J^{+}_{j,5} | \Sigma^{-} (\bm{0}) \rangle 
&=& 
\frac{1}{\sqrt{6}} 
G_{\Lambda\Sigma}(\D_{\L\S}^2) \, \,
\ol u_{\Lambda}(\bm{0})  \gamma_j \gamma_5 u_{\Sigma} (\bm{0})
,\nonumber \\
\langle \Xi^{0} (\bm{0}) | J^{+}_{\mu,5}| \Xi^{-} (\bm{0}) \rangle 
&=&  
G_{\Xi\Xi} \, \,
\ol u_{\Xi} (\bm{0}) \gamma_{\mu} \gamma_5 u_{\Xi} (\bm{0})
\label{eq:trans}
.\end{eqnarray} 
In the case of the transition matrix element,
the 
$\L$-$\S$ 
mass splitting leads to non-vanishing energy transfer,
$\D_{\L\S} = M_{\S} - M_{\L}$.
We have taken the 
$j$-th
spatial component of the 
$\L$-$\S$ 
axial current to eliminate the contribution from the pseudoscalar form factor.
The axial charge of the transition is defined by the axial form factor at vanishing four-momentum transfer,
$G_{\L \S} = G_{\Lambda\Sigma}(0)$.
There are additional non-vanishing hyperon matrix elements of 
$J^+_{\mu,5}$,
but these are related to those given in Eq.~\eqref{eq:trans} by isospin.
Our normalization conventions are those in~\cite{Jiang:2008aqa}.

To compute the axial charges 
$G_{\S\S}$, 
$G_{\L\S}$, 
and
$G_{\X\X}$, 
we utilize two-flavor \CPT\ for hyperons. 
The Lagrangian for the 
$S=1$, 
and 
$S=2$ 
hyperons has been formulated in~\cite{Tiburzi:2008bk},
and we use the isospin multiplets of that work.
At leading-order in the chiral and heavy baryon expansions, 
the axial current is matched onto the operator
\begin{eqnarray} \label{eq:current}
J_{\mu,5}^+
&=&
\sqrt{\frac{2}{3}}
g_{\L \S} 
\left[ 
\tr \left( \ol \S S_\mu \tau^+ \right) \L
+ 
\ol \L 
\, \tr \left( S_\mu \tau^+ \S \right)
\right]
\notag \\
&& +
g_{\S\S}  \, \tr \left( \ol \S [ \tau^+, \S ] \right)
+
2 g_{\X\X} 
\left( \ol \X S_\mu \tau^+ \X \right)
.\end{eqnarray}
We have omitted a surface term for the 
$\L$-$\S$ 
transition current~\cite{Beane:2003yx},
as it only contributes to the pseudoscalar form factor. 
The leading-order current produces the axial charges:
$G_{\S \S} = g_{\S \S}$, 
$G_{\L \S} = g_{\L \S}$, 
and
$G_{\X \X} = g_{\X \X}$. 
Notice we employ lower-case letters for the chiral limit values of the axial couplings,
and upper-case letters for the axial charges. 
At leading order, the two are identical.

Beyond leading order,
there are chiral corrections to the axial currents arising from two sources:
local terms and long-distance pion loops. 
The local corrections arise from the NLO axial current operator
that depends on unknown low-energy constants that we label
$A_{\S \S}$, 
$A_{\L\S}$, 
and 
$A_{\X\X}$. 
In the isospin limit, 
this operator has a form
identical to that in Eq.~\eqref{eq:current}
with the replacement
$g_{BB'} \to A_{BB'} m_\pi^2 / \L_\chi^2$, 
where 
$\L_\chi$ 
is the chiral symmetry breaking scale, 
$\L_\chi = 2 \sqrt{2} \pi f$, 
and 
$f = 132 \, \texttt{MeV}$
is the pion decay constant.
The unknown parameters are expected to be of natural size.
Pion loop contributions give rise to the long-distance corrections
to axial current matrix elements. 
These loops are generated from vertices in the hyperon chiral Lagrangian; 
and,
at this order, 
the pion-hyperon coupling constants are just the chiral limit axial couplings.
Additional non-analytic dependence on the pion mass arises from 
including the nearby spin-$\frac{3}{2}$ resonances. 
These contributions are important in light of lattice applications because 
the lattice pion masses are not considerably smaller than the 
spin-$\frac{3}{2}$--spin-$\frac{1}{2}$ 
mass splittings. 
The requisite one-loop diagrams generated from 
spin-$\frac{1}{2}$ 
and 
spin-$\frac{3}{2}$ 
pion-hyperon interactions are depicted in~\cite{Jiang:2008aqa}.  
As discussed in~\cite{Tiburzi:2009ab}, 
there is potentially a problem with including the 
$\S^*$ 
due to the kaon-nucleon threshold. 
Investigation of the relevant 
$SU(2)$ 
expansion parameter, 
however,
suggests that the threshold can be adequately described by terms analytic 
in the pion mass squared, but non-analytic in the strange quark mass.  
Thus we include the $\S^*$ baryons.

\begin{widetext}
Combining results of the tree-level and one-loop 
computations at NLO, 
we arrive at the following expressions for the hyperon axial charges
\begin{eqnarray}
G_{\Sigma\Sigma} 
&=& 
g_{\Sigma\Sigma}
+
\frac{1}{\L_\chi^2} 
\Big[ 
A_{\S \S} (\mu) m_\pi^2
-
(7 g^3_{\Sigma\Sigma}
+
4 g_{\Sigma\Sigma} )
{\cal J}(0,\mu)
+
\frac{2}{3} 
g_{\Sigma\Sigma}
g^2_{\Lambda\Sigma}
{\cal  K}(-\Delta_{\Lambda\Sigma},\mu)
+
\frac{8}{3}
g_{\Sigma\Sigma}
g^2_{\Sigma^*\Sigma}
{\cal K}(\Delta_{\Sigma\Sigma^*},\mu)
\nonumber \\
&& 
-
g_{\Sigma\Sigma}
g^2_{\Lambda\Sigma}
{\cal J}(-\Delta_{\Lambda\Sigma},\mu) 
-
\frac{8}{3}
\sqrt{\frac{2}{3}}
g_{\Sigma^*\Sigma}
g_{\Lambda\Sigma}
g_{\Lambda\Sigma^*}
\cI(-\Delta_{\Lambda\Sigma},\Delta_{\Sigma\Sigma^*},\mu)
-
\Big(
\frac{10}{9} 
g_{\Sigma^* \Sigma^*}
+
4 g_{\Sigma\Sigma}
\Big)
g^2_{\Sigma^*\Sigma}
{\cal  J}(\Delta_{\Sigma\Sigma^*},\mu)
\Big] \label{eq:GSS}
,\\
G_{\Lambda\Sigma}
&=&
g_{\Lambda\Sigma}
+
\frac{1}{\L_\chi^2}
\Big[
A_{\L \S}(\mu) m_\pi^2
-
4 
g_{\Lambda\Sigma}
{\cal  J}(0,\mu)
-
6 
g_{\Lambda\Sigma}
g^2_{\Lambda\Sigma^*}
{\cal  J}(\Delta_{\Lambda\Sigma^*}, \mu)
+
2 g_{\Lambda\Sigma}
g^2_{\Sigma\Sigma}
{\cal K}(\Delta_{\Lambda\Sigma},\mu)
-
3
g_{\Lambda\Sigma}
g^2_{\Sigma\Sigma}
{\cal J}(0,\mu)
\notag \\
&&
-\frac{1}{3}g^3_{\Lambda\Sigma}
\cI(\Delta_{\Lambda\Sigma},-\Delta_{\Lambda\Sigma}, \mu)
-
8
\sqrt{\frac{2}{3}}
g_{\Sigma\Sigma}
g_{\Sigma^*\Sigma}
g_{\Lambda\Sigma^*}
{\cal K}(\Delta_{\Lambda\Sigma^{\star}}, \mu)
+
\frac{20}{3}
\sqrt{\frac{2}{3}}
g_{\Sigma^*\Sigma^*}
g_{\Sigma^*\Sigma}
g_{\Lambda\Sigma^*}
\cI(\Delta_{\Lambda\Sigma^*},\Delta_{\Sigma\Sigma^*},\mu)
\nonumber \\
&&
-
2
g_{\Lambda\Sigma}
g^2_{\Sigma^*\Sigma}
{\cal  J}(\Delta_{\Sigma\Sigma^*},\mu)
-
\frac{3}{2}
g^3_{\Lambda\Sigma}
{\cal J}(\Delta_{\Lambda\Sigma},\mu)
-
\frac{1}{2}
g^3_{\Lambda\Sigma}
{\cal J}(-\Delta_{\Lambda\Sigma},\mu)
+
\frac{8}{3}
g_{\Lambda\Sigma}
g^2_{\Lambda\Sigma^*}
\cI(\Delta_{\Lambda\Sigma^*},-\Delta_{\Delta\Sigma},\mu)
\notag \\
&& 
-
\frac{8}{3}
g_{\Lambda\Sigma}
g^2_{\Sigma^*\Sigma}
\cI(\Delta_{\Lambda\Sigma},\Delta_{\Sigma\Sigma^*},\mu)
\Big]
, \quad \text{and} \label{eq:GLS}\\
G_{\Xi\Xi} 
&=& 
g_{\Xi\Xi}
+
\frac{1}{\L_\chi^2} 
\Big[
A_{\X \X}(\mu) m_\pi^2
 -
4
(
2g^3_{\Xi\Xi}
+
g_{\Xi\Xi})
{\cal  J}(0,\mu) 
- 
\Big(
6
g_{\Xi\Xi} 
+ 
\frac{10}{9}
g_{\Xi^*\Xi^*}
\Big)
g^2_{\Xi^*\Xi}
{\cal  J}(\Delta_{\Xi\Xi^*},\mu)
- 
\frac{8}{3}
g_{\Xi\Xi}
g^2_{\Xi^*\Xi}
{\cal   K}(\Delta_{\Xi\Xi^*},\mu)
\Big]
.\notag \\
\label{eq:GXX}
\end{eqnarray}
The nonanalytic functions appearing above can be expressed in terms of a function 
$\cF(m_\pi, \d, \mu)$, 
namely
\begin{eqnarray}
\cI(\d_1,\d_2,\mu) 
&=& 
-\frac{2}{3}\frac{1}{\d_1 - \d_2}
[\cF(m_\pi,\d_1,\mu)
-
\cF(m_\pi,\d_2,\mu)
], \, \,
{\cal J}(\d,\mu) 
=
\cI(\d,\d,\mu), 
\, \, \text{and} \, \, 
{\cal K}(\d,\mu) 
=
\cI(\d,0,\mu) = \cI(0,\d,\mu),
\notag
\end{eqnarray}
with the pion mass dependence kept implicit. 
Any terms analytic in the pion mass-squared are subsumed into the local contribution. 
Furthermore we add to the Lagrangian the 
appropriate terms to preserve the chiral limit values of the 
axial couplings. 
Accordingly the loop corrections vanish in the chiral limit, 
$\cF(0,\d,\mu) = 0$,
see e.g.~\cite{Tiburzi:2005na} 
for the 
explicit expression for 
$\cF(m,\d,\mu)$. 
The various couplings and mass splittings that also appear in the NLO expressions will be discussed below. 
\end{widetext}


{\bf Pion mass dependence}.
We use both physical and lattice QCD inputs to explore the pion mass dependence
of the hyperon axial charges in two-flavor \CPT. 
Ideally we would use lattice data to study the pion mass dependence of the
axial charges and deduce the unknown couplings (and potentially indirectly verify values of known couplings). 
Armed with knowledge of these parameters, 
we could diagnose the behavior of $SU(2)$ expansions in each strangeness sector, 
extrapolate the lattice data to the physical pion mass, \emph{etc}. 
The first calculation of hyperon axial charges~\cite{Lin:2007ap} 
is fortunately fully dynamical so that contact with QCD can in principle be made. 
The study is limited, however, notably by two conditions:
the pion masses lie in the range 
$350 \, \texttt{MeV} \lesssim m_\pi \lesssim 750 \, \texttt{MeV}$,
and a single lattice spacing was used in a hybrid lattice action employing different fermion 
discretizations for valence and sea quarks. 
The former condition only provides us with data at the lowest pion mass 
(possibly lowest two masses)
for chiral extrapolations.%
\footnote{
For the nucleon axial charge, 
chiral extrapolations have been carried out including pion masses basically in the same range~\cite{Hemmert:2003cb,Procura:2006gq}.
We have carried out similar extrapolations for the hyperons. 
While high quality fits to the pion mass dependence of nucleon and hyperon axial charges result, 
one should question the remarkable ability of the effective theory to work outside the range of its applicability.
Estimating the size of neglected higher-order terms casts doubt as to whether the effective theory is
under control for the range of pion masses used. 
Such error estimates have been made for nucleon mass extrapolations%
~\cite{Beane:2004ks}. 
}  
The latter further complicates extrapolation
because valence-sea meson masses are additively renormalized
by a non-negligible amount~\cite{Bunton:2006va,Orginos:2007tw,Aubin:2008wk}, 
necessitating a partially quenched treatment.%
\footnote{ 
Partially quenched computations exist for the case of three light valence quark flavors~\cite{Detmold:2005pt}.
}
Because chiral extrapolation of the data in~\cite{Lin:2007ap} would be premature, 
we pursue a more modest goal of estimating the pion mass 
dependence of the axial charges using input from lattice QCD.

Table~\ref{t:table}
summarizes input used to study hyperon axial charges. 
Additionally we compare with the case of the nucleon axial charge,
$G_A$.%
\footnote{
For reference, the nucleon axial charge at NLO is given by~\cite{Bernard:1998gv}
\begin{eqnarray}
G_{A} 
&=& 
g_{A} 
+ 
\frac{1}{\L_\chi^2} 
\Big[
A_{NN}(\mu) m_\pi^2
-
4
(2g^3_A +g_A)
{\cal J}(0,\mu)
\nonumber \\
&&
+
\frac{64}{9} g_A g^2_{\Delta N} 
{\cal K}(\Delta,\mu)
-8 \Big(g_A + \frac{25}{81} g_{\Delta\Delta} \Big) g^2_{\Delta N}
{\cal J}(\Delta,\mu)\Big]
\notag \label{eq:GNN}
.\end{eqnarray}
}
%
%
%
%
%
%
%
%
%
%
%
%
%
\begin{table}[t]
\begin{center}
\begin{tabular}{cc|c}
Input Parameters &  & $\quad$ Source $\quad$
\tabularnewline
\hline
\hline
$\D = 290 \, \texttt{MeV}$ &
$\D_{\L\S} = 77 \, \texttt{MeV}$ &
Expt. 
\tabularnewline
$\D_{\L\S^*} = 270 \, \texttt{MeV}$ &
$\D_{\X\X^*} = 215 \, \texttt{MeV}$ & 
\tabularnewline
\hline
$G_A (139 \, \texttt{MeV} )= 1.27$ &
$G_{\L \S} (139 \, \texttt{MeV} ) = 1.47$ &
Expt.
\tabularnewline
$g_{\D N} = 1.48$ &
$g_{\L \S^*} = -0.91$&
\tabularnewline
$g_{\S^*\S} = 0.76$ &
$g_{\X^* \X} = 0.69$ &
\tabularnewline
\hline
\multicolumn{2}{c|}{$g_{\D \D} = - 2.2 \quad g_{\S^*\S^*} = -1.47 \quad g_{\X^* \X^*} = -0.73$}
&
$SU(3)$~\cite{Butler:1992pn}
\tabularnewline 
\hline
$G_{\S\S} (139 \, \texttt{MeV} ) = 0.78$ &
$G_{\X\X} (139 \, \texttt{MeV} ) = 0.24$ &
Extrap.~\cite{Lin:2007ap} 
\tabularnewline
\hline
\hline
\tabularnewline
\end{tabular}
\begin{tabular}{cc}
Output Parameter Estimates
\tabularnewline
\hline
\hline
$g_A = 1.18$ & $A_{NN}(\L_\chi) = -12.0$ 
\tabularnewline
$g_{\S\S} = 0.73$ & $A_{\S\S}(\L_\chi) = -2.9$ 
\tabularnewline
$g_{\L \S} = 1.29$ & $-$
\tabularnewline
$g_{\X \X} = 0.23$ & $A_{\X\X}(\L_\chi) = -0.22$
\tabularnewline
\hline
\hline
\end{tabular}
\end{center}
\caption{%
Summary of input and output parameters. 
Cited lattice values at the physical pion mass 
are obtained from an empirical extrapolation unrelated to \CPT\ 
that describes the data remarkably well. 
Without lattice data for the 
$\L$-$\S$ 
transition, 
we assume 
$A_{\L \S} (\L_\chi) \equiv 0$.
Estimated values for output parameters 
result from using the listed input parameters
as well as lattice data at the lightest valence pion mass, 
$m_\pi = 354 \, \texttt{MeV}$~\cite{Lin:2007ap}.
To use this pion mass, 
we are neglecting the additive mass renormalization
of valence-sea mesons due to discretization.
}
\label{t:table}
\end{table}
Mass splittings are taken from experiment, 
as are the known axial charges, 
and 
spin-$\frac{3}{2}$-to-spin$\frac{1}{2}$ 
axial transition couplings. 
For the spin-$\frac{3}{2}$ axial charges, 
we assume 
$SU(3)$
symmetry, 
and use input values that are coincidentally not too different from 
$SU(6)$ 
quark model predictions. 
The remaining unknown parameters are estimated using 
results from lattice QCD.

\begin{figure}
\begin{center}
\includegraphics[width=0.3\textwidth]{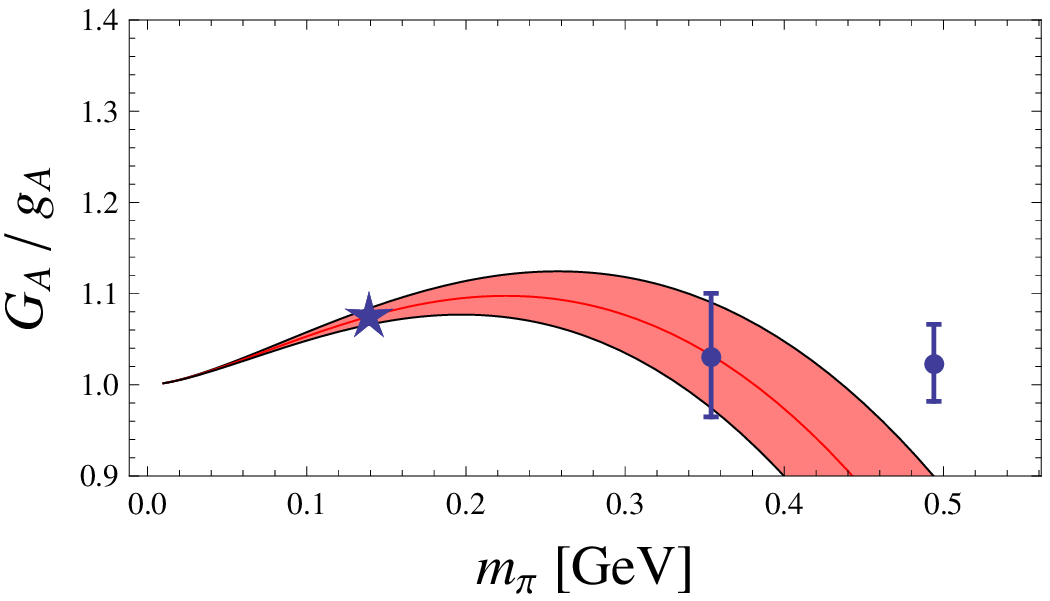}
\includegraphics[width=0.3\textwidth]{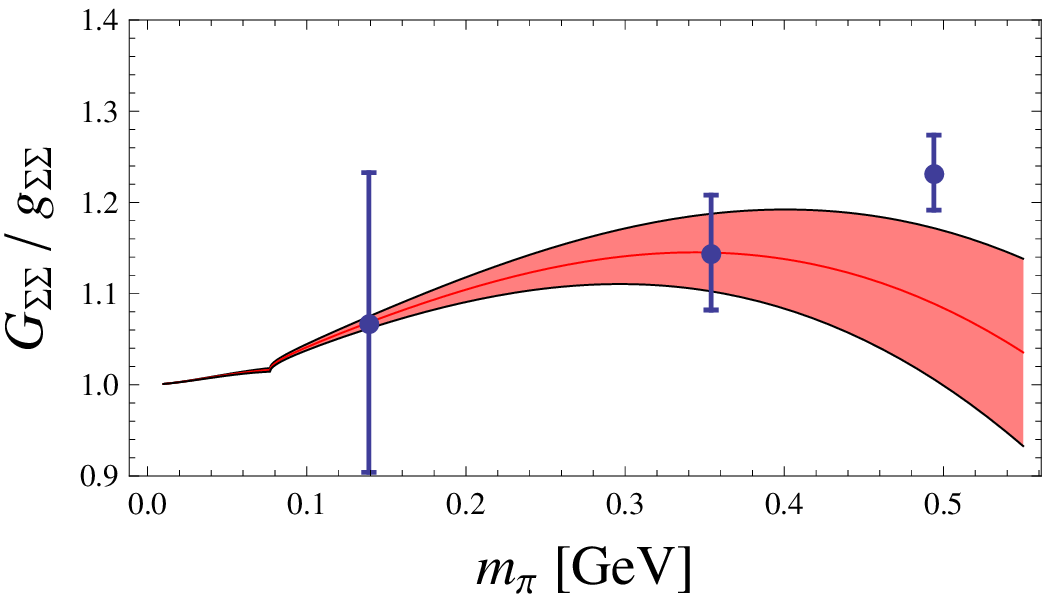}
\includegraphics[width=0.3\textwidth]{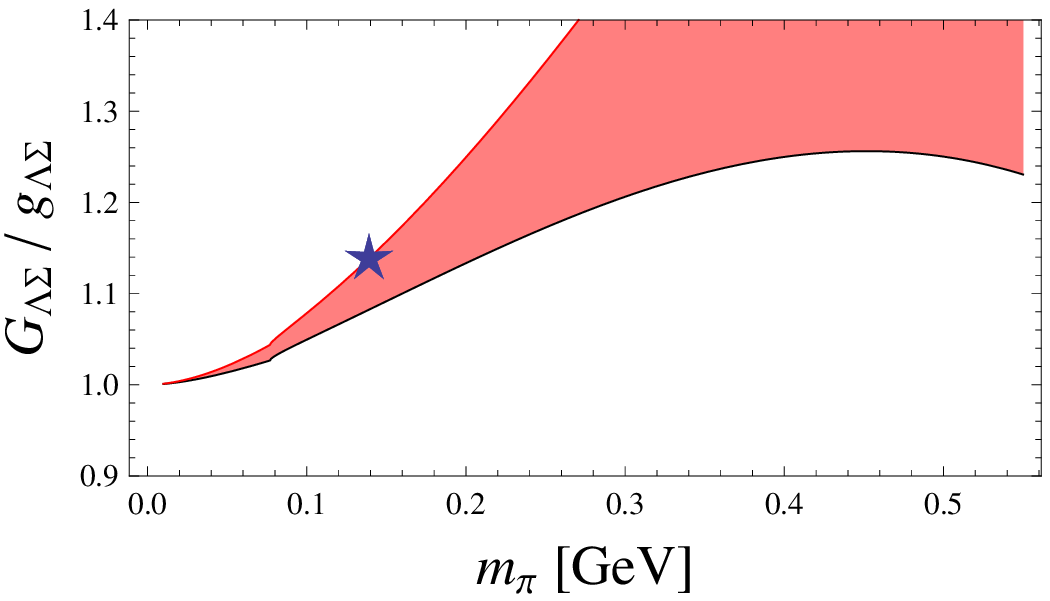}
\includegraphics[width=0.3\textwidth]{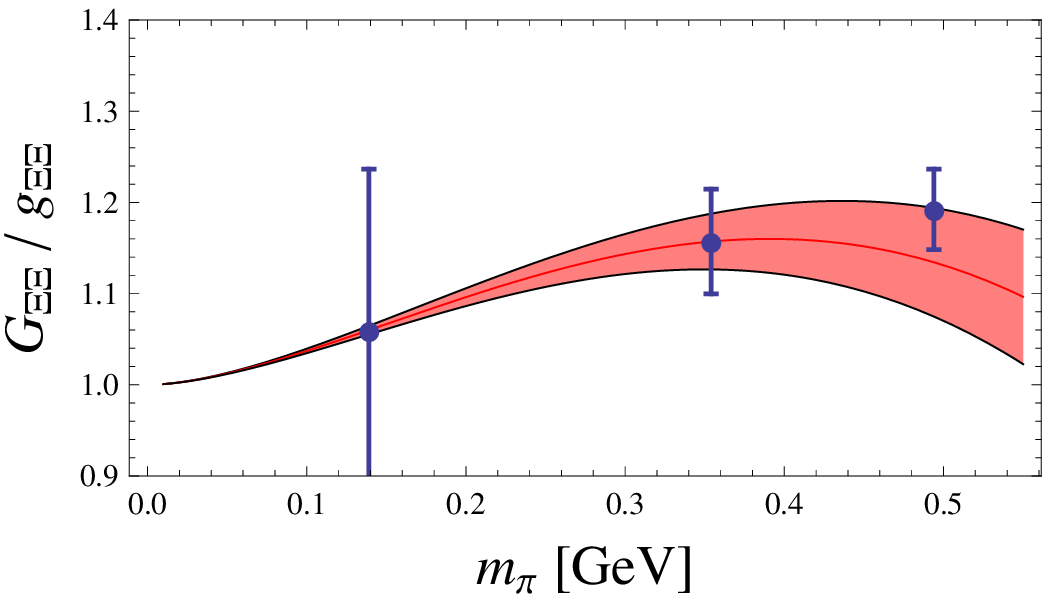}
\end{center}
\caption{Pion mass dependence of isovector axial charges. 
Stars denote physical inputs, 
while the points are lattice QCD results taken from \cite{Lin:2007ap},
of which the value at the physical pion mass is obtained from an empirical quark-mass extrapolation,
and the lowest mass data are used to estimate chiral limit couplings and local contributions at NLO. 
For 
$G_{\L\S}$ 
and 
$G_{\S\S}$, 
cusps appear at
$m_\pi = \D_{\S\L}$ 
where the $\Sigma$ becomes unstable to strong decay. 
Below this value, 
we plot the real part; 
the imaginary part is negative and an order of magnitude smaller. 
}
\label{chiralcorrection}
\end{figure}

With the values obtained, 
we plot the pion mass dependence of the various axial charges in Fig.~\ref{chiralcorrection}, 
where ratios of each axial charge to its chiral limit value are plotted. 
As the ratios are NLO values relative to their LO values,
they thereby give an indication of when the expansion is perturbative. 
Because we set the NLO local contribution for the 
$\L$-$\S$ 
transition to zero, 
we vary it,
$- 4 < A_{\L \S}(\L_\chi) < 0$, 
with the sign chosen to be consistent with the NLO contributions of the other baryons. 
To obtain an estimation of the errors resulting from neglected higher-order terms for the remaining baryons, 
we construct a band corresponding to varying local corrections arising at
next-to-next-to-leading order in the chiral and heavy baryon expansions,
which have the form
\begin{equation}
\d G_{BB}
=
A'_{BB} \, g_{BB} \frac{\D_{BB^*}  \, m_\pi^2}{ M_B \,  \L_\chi^2}
.\end{equation}
The bands arise from the range of values, 
$-2 < A'_{BB} < 2$.

For the nucleon,  
the estimated local term in Table~\ref{t:table} is unnaturally large in magnitude,
and the curve in Fig.~\ref{chiralcorrection} does not follow the lattice data.
This suggests that higher-order terms are important, 
or the value of 
$g_{\D\D}$ 
may be different---this coupling is also not likely under control in \CPT~\cite{Jiang:2008we}.  
For the $\S$, 
the estimated local term is four times smaller in magnitude, and
the curve better follows the trend in the lattice data. 
For the $\X$, 
the local term is even smaller in magnitude, 
and the trend in the lattice data is even better matched. 
We conclude that the empirical quark mass extrapolation
performed in~\cite{Lin:2007ap} is likely consistent
with an 
$SU(2)$ 
analysis of the 
$\S$ 
and 
$\X$ axial charges. 
Furthermore, 
it appears that the 
$S=2$ 
sector has the best convergence properties. 
To verify these claims, 
lattice data at smaller pion masses are needed. 
Additionally lattice calculation of the 
$\L$-$\S$ 
axial transition will better determine parameters, 
as the 
$\L$ 
and 
$\S$ 
systems are coupled.     
Extraction of 
$G_{\L \S}$ 
can best be performed with isospin twisted boundary conditions~\cite{Tiburzi:2005hg}.

\smallskip

\begin{acknowledgments}												   %
This work is supported in part by the 
U.S.~Dept.~of Energy, Grant No.~DE-FG02-93ER-40762 (B.C.T.), 
and by the Schweizerischer Nationalfonds (F.-J.J.). 
\end{acknowledgments}												   %

%
\bibliography{hb}

\end{document}